\documentstyle[12pt,epsf,psfig]{article}

\newcommand{\be}{\begin{equation}}
\newcommand{\ee}{\end{equation}}
\newcommand{\msbox}[1]{\mbox{\scriptsize #1}}
\newcommand{\mx}{_{\msbox{max}}}

\begin{document}
\title{\bf Is DLA Locally Isotropic or Self-Affine?}
\author{{\Large Rainer Hegger and Peter Grassberger}\\[2ex]
Physics Department,
University of Wuppertal\\
D - 42097 Wuppertal, Germany}
\maketitle

\begin{abstract}
We present results of simulations which show unambiguously
that DLA clusters are not self-affine, in contrast to frequent
claims. The measured observable is the asymmetry of the last step
of a walker before he sticks to the growing cluster. Using
deposition onto an originally straight line off lattice, we show
that this asymmetry tends to zero algebraically
with the thickness of the deposit.
\end{abstract}

\newpage

Though diffusion limited aggregation (DLA) \cite{witt-sand} has
been studied very intensively during the last decade
\cite{meak}, its theoretical understanding
remains still very unsatisfactory. There have been a number of
isolated results such as Halsey's relations for the generalized
dimensions of the growth measure in 2-$d$ \cite{halsey}, or the
results of \cite{eckmann} on the shape
of large clusters on 2-$d$ regular lattices, but very basic
properties of DLA clusters are still unclear.

One such basic question is whether DLA clusters themselves (not
the growth measure on them!) are multifractal \cite{vic-fam-meak},
become space-filling at infinite size \cite{mandelbr} or
at least show multiscaling \cite{conig,oss}.

Even more basic is the
question whether DLA clusters are isotropic on small scales. On
large scales it is of course clear that they are not.
Rather, the direction towards the center of the cluster is clearly
singled out. But it might well be that this anisotropy is lost as
one goes into further and further ramifications of the main branches
of the cluster. By
looking at their (rather small) clusters, the authors of the first
paper on DLA \cite{witt-sand} claimed that DLA should indeed be
isotropic
and self-similar on small scales. But this was questioned
by a more systematic study \cite{meak-vic}, in which a strong
anisotropy was found.

Unfortunately, the analysis of \cite{meak-vic} is marred by a
number of problems. It was done on a square lattice which
is known today to lead to serious artifacts
\cite{vicsek}, a parametrization of the two-point
correlation function was used which is neither scale-invariant
nor self-affine, and no systematic extrapolation towards
the limit of infinite clusters was attempted. Finally,
working in a radial geometry as in \cite{meak-vic} induces
rather awkward finite-size effects.

The latter can be avoided by using not a single site as a seed of the
cluster, but an infinite line (i.e., a finite line with periodic
boundary conditions). This is often called diffusion limited {\it
deposition} (DLD), in order to distinguish it from DLA proper. In DLD,
the anisotropy on large scales is even more clearly seen than in DLA,
and its origin is well understood.  It is mostly on the basis of this
large scale anisotropy that it is often claimed \cite{vicsek} that DLA
is self-affine.
\footnote{There exists one study \cite{meak-kert-vic}
which finds very strong anisotropy effects in DLD also on small
scales, but this was done on a square lattice with
strong noise reduction. It is almost sure that
the results of \cite{meak-kert-vic} are lattice artifacts.}

In the following we shall present numerical data obtained in
off-lattice simulations which clearly
show that DLD becomes isotropic in the limit
of infinite thickness of the deposit, on scales much smaller than
this thickness.
Since its microscopic structure should be the same as in DLA,
this shows that also DLA should become isotropic on small scales.

Before proceeding, we should make a few comments on self-affinity.
A set is self-affine (in the statistical sense), if it (resp. its
statistical characteristics) is invariant under an {\it affine},
i.e. inhomogeneous linear transformation which is not just a
isotropic rescaling
plus a translation. The prototype of a
self-affine critical phenomenon is directed percolation.
There, the anisotropy is seen as soon as a scale larger
than one lattice unit is considered. There is no intermediate
range of scales in which lattice effects disappear, but anisotropy
is not yet fully pronounced. It is precisely the existence of such
a range which is claimed in the present note.

We study DLD off lattice in 2 dimensions, with periodic boundary
conditions (period $L$) sidewise. The
step length in the diffusion process is variable, depending on the
distance from the cluster. Particles are presented by discs of
radius $R$. If the distance between the $n$-th diffusing
particle and the nearest particle on the cluster (resp. to the flat
initial surface) is $r$, the length of the next step is chosen as
$l\le r+\epsilon R$, with $\epsilon=0.05$.
Since $\epsilon>0$, such a step can lead to a overlap with the
cluster. If this happens, the steps length is reduced such that
the particle just touches the cluster, the cluster is updated by
incorporating the particle, and the angle $\theta_n$ of the last
step (relative to the normal to the base line) is recorded
(fig.1).

\begin{figure}[htb]
\caption{Schematic drawing of a particle approaching and sticking to
a deposit during its very early stage.}
\label{fig.1}
\end{figure}

In fig.2 we show average values of $\cos \theta_n$ against $n$.
The lateral size of the system was chosen as $L=5120R$, and
$n$ went up to $n\mx=4\times10^5$. If we
assume that the density in the deposit decays as $h^{-.3}$
with height $h$ \cite{meak-fam}, this implies that the
deposit grows to an average thickness $ h\mx\approx
R(n\mx R/L)^{1.43} \approx 10^3R$. This was indeed
observed, and it means that all our simulations correspond to
the limit $h\mx << L$ in which the system can be regarded as
having an essentially infinite lateral extension, and we do
not have to worry about finite-$L$ corrections. The data shown
in fig.2 represent the results of 50 such runs.

\begin{figure}[htb]
\caption{Log-log plot of $\langle \cos\theta\rangle$, where
$\theta$ is the angle of a particle's velocity just before it
sticks, against $nR/L$. In order to reduce statistical fluctuations,
data are binned into intervals with $\Delta n/n =0.02$.}
\label{fig.2}
\end{figure}

Apart from rather strong corrections at small $n$ (which are
easily explained from the fact that the initial flat surface
is very different in its microstructure from the later surfaces
consisting of small discs), we see a very clear scaling behavior.
More precisely, we find
\be
   \langle \cos \theta \rangle \sim n^{-\alpha}
\ee
with
\be
   \alpha = 0.23\pm0.03
\ee

This scaling law seems unrelated to any previously observed
scaling law in DLA, and by itself should be interesting. But
more interesting is that it demonstrates quite convincingly that
$\lim_{n\to\infty}
\langle \cos \theta \rangle = 0$, i.e. that particles approach
the cluster {\it isotropically} before the stick. On the other
hand, the power $\alpha$ is not very large, i.e. the convergence
of $\langle \cos \theta \rangle$ is very slow. This explains why
analyses where no extrapolation towards $n=\infty$ is made
\cite{meak-vic} should find strong anisotropies.

Let us now discuss our result in view of other results about DLA.
First of all, we stress again that it does not mean that DLA or DLD
is isotropic on large scales ($\geq L$). The reason why it cannot be
so is well understood.

Secondly, there exists a result which at first seems to contradict our
conclusions. That is the finding of \cite{meak-maj,oss2} that shortest
paths in DLA (and presumably also in DLD) have dimension 1. Thus the
stems of typical large branches are straight, which seems to be in
conflict with our claim that typical branches grow non-straight. The
resolution of this conflict is that straight branches will grow longer
than curled ones, shield them from further growth, and will finally
form the stems of large arms. Thus it is posterior selection which is
responsible for the straightness, not the fact that all branches are
straight from the outset.

Finally, our claim is fully compatible with the result of
\cite{meak-fam} that in the limit $h>>L^{\nu_\perp}$ the deposit has
a surface of thickness $\sim L$, i.e. that the structures in this surface
are roughly isotropic --- which also shows that the deposit is not
self-affine.

\vspace{.3cm}

One of us (P.G.) wants to thank L. Pietronero for a most interesting
discussion, and Antonio Politi for the warm hospitality at the Istituto
Nazionale di Ottica, Firenze, where some of the work was done.

\end{document}